# High-resolution diffusion-weighted imaging at 7 Tesla: single-shot readout trajectories and their impact on signal-to-noise ratio, spatial resolution and accuracy

Sajjad Feizollah* [a, b], Christine L. Tardif [a, b, c]

a. Department of Neurology and Neurosurgery, Faculty of Medicine and Health Sciences, McGill University, 3801 Rue University, Montreal, QC, Canada

b. McConnell Brain Imaging Centre, Montreal Neurological Institute, McGill University, 3801 Rue University, Montreal, QC, Canada

c. Department of Biomedical Engineering, Faculty of Medicine and Health Sciences, McGill University, Duff Medical Building, 3775 Rue University, Suite 316, Montreal, QC, Canada

* Corresponding author: Sajjad Feizollah: sajjad.feizollah@mail.mcgill.ca

Christine Lucas Tardif: christine.tardif@mcgill.ca

**Highlights:**

- Investigated trade-off between effective resolution and SNR for dMRI with EPI, partial Fourier EPI and spiral trajectories at 7T.
- NMR field probes were used to minimize artifacts due to eddy currents and field non-uniformities.
- For matching nominal resolutions, EPI has the highest effective resolution, specificity, and sharpening measures due to T2* k-space modulation.
- For matching effective resolutions, spirals have the highest SNR making them the optimal readout trajectory for high-resolution dMRI of the brain at 7T.

## Abstract

Diffusion MRI (dMRI) is a valuable imaging technique to study the connectivity and microstructure of the brain in vivo. However, the resolution of dMRI is limited by the low signal-to-noise ratio (SNR) of this technique. Various multi-shot acquisition strategies have been developed to achieve sub-millimeter resolution, but they require long scan times which can be restricting for patient scans. Alternatively, the SNR of single-shot acquisitions can be increased by using a spiral readout trajectory to minimize the sequence echo time. Imaging at ultra-high fields (UHF) could further increase the SNR of single-shot dMRI; however, the shorter T2* of brain tissue and the greater field non-uniformities at UHFs will degrade image quality, causing image blurring, distortions, and signal loss.

In this study, we investigated the trade-off between the SNR and resolution of different k-space trajectories, including echo planar imaging (EPI), partial Fourier EPI, and spiral trajectories, over a range of dMRI resolutions at 7T. The effective resolution, spatial specificity and sharpening effect were measured from the point spread function (PSF) of the simulated diffusion sequences for a nominal resolution range of 0.6-1.8 mm. In-vivo partial brain scans at a nominal resolution of 1.5 mm isotropic were acquired using the three readout trajectories to validate the simulation results. Field probes were used to measure dynamic magnetic fields offline up to the $3^{rd}$ order of spherical harmonics. Image reconstruction was performed using static $\Delta B_0$ field maps and the measured trajectories to correct image distortions and artifacts, leaving T2* effects as the primary source of blurring. The effective resolution was examined in fractional anisotropy (FA) maps calculated from a multi-shell dataset with b-values of 300, 1000, and 2000 s/mm$^2$ in 5, 16, and 48 directions, respectively. In-vivo scans at nominal resolutions of 1, 1.2, and 1.5 mm were acquired and the SNR of the different trajectories calculated using the multiple replica method to investigate the SNR. Finally, in-vivo whole brain scans with an effective resolution of 1.5 mm isotropic were acquired to explore the SNR and efficiency of different trajectories at a matching effective resolution. FA and intra-cellular volume fraction (ICVF) maps calculated using neurite orientation dispersion and density imaging (NODDI) were used for the comparison. The simulations and in vivo imaging results showed that for matching nominal resolutions, EPI trajectories had the highest specificity and effective resolution with maximum image sharpening effect. However, spirals have a significantly higher SNR, in particular at higher resolutions and even when the effective image resolutions are matched. Overall, this work shows that the higher SNR of single-shot spiral trajectories at 7T allows us to achieve higher effective resolutions compared to EPI and PF-EPI to map the microstructure and connectivity of small brain structures.

**Keywords:**

Diffusion MRI, ultra-high field, field monitoring, T2* blurring, effective resolution, signal-to-noise ratio.

# 1 Introduction

Diffusion MRI (dMRI) is sensitive to the motion of water molecules in tissue and thus provides insight into its microstructure (Afzali et al., 2021; Basser et al., 1994; Jones, 2010). As gradient pulses are employed to encode diffusion in a specific direction, the loss of phase coherence due to motion along that direction results in attenuation of the MR signal (Jones, 2010; Tanner, 1979). This signal attenuation together with long diffusion-encoding times significantly reduces the signal-to-noise ratio (SNR) of dMRI and thus limits the spatial resolution that can be achieved (Polders et al., 2011; Polzehl and Tabelow, 2016). 2-D fast imaging approaches typically used for diffusion imaging further reduce the SNR compared to 3D acquisitions normally performed for anatomical scans. Several imaging techniques have been implemented in dMRI to reduce the echo time (TE) in order to minimize the signal loss due to T2* decay without sacrificing scan time, including accelerated parallel imaging (Griswold et al., 2002; Pruessmann et al., 2001, 1999), partial Fourier echo-planar imaging (PF-EPI) (e.g., Noll et al., 1991; Blaimer et al., 2009), non-Cartesian trajectories such as spirals (Block & Frahm, 2005; Assländer et al., 2013), and high performance gradients (Foo et al., 2020; Setsompop et al., 2013; Wang et al., 2021).

Complementary techniques have also been developed to increase image resolution that are based on acquiring k-space in multiple shots. These techniques include acquiring multiple interleaves in the phase encode direction (e.g., Butts et al., 1996), multiple segments in the readout direction (e.g., Robson et al., 1997; Porter & Heidemann, 2009; Heidemann et al., 2010), using multi-shot non-Cartesian trajectories (e.g., Liu et al., 2004; Wang et al., 2005; Pipe & Zwart, 2006; Truong & Guidon, 2014), and 3-D multi-slab acquisitions (Dai et al., 2021; Engström and Skare, 2013; Moeller et al., 2020; Wu et al., 2016). These multi-shot techniques are sensitive to phase differences between shots due to motion and artifacts caused by physiological motion such as breathing, which must be corrected (e.g., Chen et al., 2013; Guhaniyogi et al., 2016; Mani et al., 2017). The above techniques can be combined with g-slider, a multi-shot technique to increase resolution along the slice direction. g-Slider uses a tailored RF pulse profile to excite a slab that modulates single slice information. The acquisition is repeated the same number of times as the slice number each with different RF pulses, and then individual slices are unaliased using the acquired scans (Setsompop et al., 2018; Ramos‐Llordén et al., 2020; Wang et al., 2021; Ramos-Llordén et al., 2022). Using this method, resolutions as high as 500 μm have recently been achieved (Liao et al., 2022). In addition to these techniques, reduced field-of-view (rFOV) imaging has been also proposed in which reduction in the FOV results in an increased distance between two adjacent k-space lines allowing shorter readout duration to minimize T2* signal decay (e.g., Feinberg et al., 1985; Karampinos et al., 2009; Saritas et al., 2014). Although this method covers a small region, it can be used repetitively for a whole-brain acquisition which increases the scan time similar to other techniques. Using these techniques, a typical diffusion-weighted sequence with 64 directions can take ~ 45-60 minutes, which limits the application of high-resolution dMRI in clinical research. This has motivated the development and optimization of single-shot readout approaches to improve dMRI SNR and resolution.

One way to boost the SNR is to scan at ultra-high magnetic field (UHF), which offers an increase in the intrinsic sensitivity and thus the opportunity to acquire high-resolution scans. The SNR has a supralinear ($\sim B_0^{1.95}$) relationship with the main magnetic field ($B_0$) over a range of about 3 to 7 T (Pohmann et al., 2016). However, due to shorter T2 and T2* relaxation times at UHFs leading to a faster signal decay, the benefit of UHF imaging for dMRI depends on the echo time (Gallichan, 2018; Uğurbil et al., 2013). Efficient readout trajectories that reduce TE can maximize the SNR increase provided by UHF imaging.

Single-shot spiral acquisitions are among the most efficient trajectories (Assländer et al., 2013; Engel et al., 2018; Lee et al., 2021a; Wilm et al., 2017). Center-out spiral trajectories minimize the echo time by starting acquisition from the k-space center, resulting in a significant SNR advantage (Lee et al., 2021a). Furthermore, acquiring with a spiral pattern avoids sharp changes in the trajectory direction that decrease speed due to limitations in gradient slew rates and peripheral nerve stimulation (PNS). Additionally, spiral trajectories inherently have zero gradient moments at the k-space center which make them robust to flow artifacts (Nishimura et al., 1995). The disadvantage of this type of k-space sampling is increased sensitivity to gradient imperfections and $B_0$ field non-uniformities that cause image blurring and ring-shaped artifacts (Block and Frahm, 2005).

The development of field monitoring probes allows us to measure dynamic field imperfections and use this information during image reconstruction to minimize image artifacts. Application of these field probes has significantly improved dMRI image quality for EPI and spiral trajectories at 3T at a nominal in-plane resolution of 1.3 mm (Lee et al., 2021a; Wilm et al., 2015, 2017), and at 0.69 mm using high-performance gradients (Wilm et al., 2020). Ma and colleagues (2020) used field monitoring probes to correct artifacts caused by gradient imperfections in the Human Connectome Project with an isotropic resolution of 1.05 mm diffusion EPI protocol at 7T. To the best of our knowledge, the advantages of single-shot spirals for dMRI at 7T has not been investigated.

Although the SNR is an important factor in limiting image resolution, it is not the only contributing factor. The T2* signal decay during the readout will cause a blurring artifact that depends on the k-space sampling pattern, such that the effective resolution is lower than the nominal resolution of the scan. This blurring effect is enhanced at UHFs due to the shorter T2* relaxation times of brain tissue: T2 and T2* is nearly halved at 7T compared to 3T (Cox and Gowland, 2010; Peters et al., 2007). This lower effective resolution reduces the benefit of moving to UHF for high-resolution dMRI. Reischauer and colleagues (2012) showed that a lower effective resolution is achieved for dMRI at 7T in comparison to 3T using an EPI readout with the same acceleration factor. Engel and colleagues (2018) showed that effective resolution of a single-shot T2*-weighted GRE spiral acquisition at 7T is approximately 1.4 times higher than the nominal resolution. The impact of EPI and spiral readout trajectories with different acquisition parameters on image quality has not been thoroughly investigated at 7T.

The aim of this study is to determine the optimal single-shot readout trajectory for high-resolution dMRI at 7T by investigating the trade-off between SNR and effective resolution of various k-space trajectories. We use simulations to characterize the sole impact of T2* decay on spatial resolution and accuracy of dMRI using a PSF analysis for EPI, PF-EPI and spiral readout trajectories. In-vivo scans corrected for eddy currents and static field nonuniformities are used to validate the simulation results, and compare the SNR of the different trajectories at matching nominal resolutions. Finally, scans with matching effective resolution were acquired to investigate the SNR and efficiency of the different trajectories.

## 2 Methods

### 2.1 Artifact and blurring correction due to imperfections in spatial encoding

There are spatio-temporal deviations from prescribed magnetic field gradients during the readout, mainly due to eddy currents and concomitant fields. Furthermore, there are subject-specific static field non-uniformities ($\Delta B_0$), and dynamic field perturbations related to subject motion and physiology such as breathing. These field deviations result in the accumulation of additional phase terms during the readout as a function of spin location in space, which causes inaccuracies in spatial encoding. These inaccuracies result in ghosting artifacts, blurring, and the appearance of unwanted signal patterns that depend on the readout trajectory used (Bernstein, 2004). In order to investigate the sole effect of T2* signal decay during the readout on the PSF, image artifacts caused by these sources must first be corrected. We measured the spatio-temporal dynamics of the magnetic field using 16 field monitoring probes (Skope MRT, Zurich, Switzerland) and acquired a static $\Delta B_0$ field map. This information was included in the image reconstruction pipeline using the expanded signal encoding model described below to minimize image artifacts. The differences in image quality between the reconstructed images acquired using different k-space trajectories are therefore primarily due to T2* signal decay during the readout.

### 2.1.1 Image reconstruction using the expanded signal model

The expanded signal model is a generalized form of the Fourier transform which is typically used for image reconstruction. Unlike the Fourier transform, the power of this method is that it can model the image acquisition using any basis function for spatial encoding, and thus can include terms to describe deviations from the prescribed linear field (Wilm et al., 2011). This approach can minimize image artifacts for cartesian and non-cartesian imaging, however its application has been limited by its significant computational requirements leading to long image reconstruction times. With recent advancements in computing hardware, it is gradually finding its way into image reconstruction pipelines.

A discretized form of the expanded signal model in time and space that accounts for gradient imperfections and $B_0$ spatial non-uniformity was implemented to reconstruct images using (1),

$$s = Em \tag{1}$$

where $s$ is a matrix of samples of the measured MR signal over time, $m$ is a matrix of the magnetization in space, and $E$ is the encoding matrix of which elements are calculated as in (2).

$$E_{\gamma,r,t} = c_\gamma(r).e^{-i\varphi(r,t)} \qquad (2)$$

$c_\gamma(r)$ is the sensitivity of coil $\gamma$ at position $r$, and $\varphi(r,t)$ is the accumulated phase of a spin at position $r$ and time $t$ according to (3).

$$\varphi(r,t) = k_0(t) + \sum_{b=1}^{L} k_b(t).h_b(r) + \Delta B_0(r).t \qquad (3)$$

$k_0(t)$ is the measured zero-th order spherical harmonic term or dynamic $\Delta B_0$ over time, $k_b(t)$ is the coefficient of the spherical harmonic basis function $b$ that is calculated from the dynamic field probes measurements during the readout, and $h_b(r)$ is the spherical harmonic basis function $b$. $L$ is the number of spherical harmonics coefficients, and $\Delta B_0(r)$ is the inhomogeneity of the main magnetic field ($B_0$) at position $r$. Images are reconstructed by solving for $m$ in (1) using the Conjugate-Gradient (CG) method.

CG is an iterative reconstruction method that requires a termination criterion that is typically determined empirically. In every iteration, CG adds a small amount of noise to the solution; therefore, finding the optimal stopping point to achieve a high-quality reconstruction while avoiding excessive addition of noise is important. We used the same approach to stop the reconstruction as used by Lee and colleagues (2021b). Iteration was stopped when the difference images of two consecutive iterations had no visible structures. A minimum of 6 iterations was used. In general, higher resolutions and under-sampling factors required more iterations (up to 16). Spirals usually converged faster than EPI and PF-EPI for a given resolution.

In-house MATLAB code optimized for GPU processing was developed for image reconstruction on a workstation with Intel 11700F CPU, 64 GB of RAM, and an NVIDIA GeForce RTX 3090 graphics card with a reconstruction time of 1.8-0.3 seconds per slice, depending on the matrix size and trajectory duration.

## 2.2 Simulations

### 2.2.1 Sequence simulations

Diffusion-weighted spin-echo sequences with EPI, PF-EPI, and spiral readout trajectories were simulated in MATLAB. The excitation and refocusing pulse durations used in the simulations and in-vivo scans were set to 2.56 and 6.40 ms respectively to suppress the fat signal using the method by Ivanov et al. (2010) as used in the Human Connectome Project (Vu et al., 2015). The diffusion-encoding duration was calculated based on trajectory specifications for a b-value of 2000 s/mm$^2$ with a maximum gradient amplitude and slew rate of 73 mT/m and 200 T/m/s respectively, as used on the Siemens Terra 7T scanner.

Readout trajectories were simulated for resolutions of 0.6 to 1.8 mm isotropic with 0.1-mm increments. Fixed parameters for all trajectories include: field-of-view (FOV) = 256×256 mm$^2$, repetition time (TR) = 5000 ms, and sampling rate of 1 MHz. EPI trajectories were generated with the same gradient limitations used for the diffusion-encoding, and the following parameters: acceleration factors (R) along phase encode (PE) direction = 2, 3, and 4, bandwidth-per-pixel = 1384 Hz, PF factor = 0.75, and spatial encoding in the anterior-posterior direction. Spiral trajectories were generated using the method in (Hargreaves, 2001) with a maximum gradient amplitude of 27 mT/m and slew rate of 160 T/m/s to avoid PNS and critical acoustic resonance frequencies of the gradient system. Three spiral trajectories were generated corresponding to acceleration factors R of 4, 5, and 6 respectively.

### 2.2.2 Point spread function characterisation

For the PSF analysis, a single point in the center of the image domain was simulated with T1, T2, and T2* relaxation times of the GM/WM set to 1300/800, 72/79, 66/46 ms at 3T, and 2000/1200, 47/47, 33/26 ms at 7T respectively (Cox and Gowland, 2010; Peters et al., 2007; Rooney et al., 2007; Wansapura et al., 1999). The simulated signal decay was sampled at the time points along the different trajectories to fill k-space. For PF-EPI, the missing part of k-space was filled based on the conjugate symmetry feature of k-space.

In EPI-based trajectories, considerable signal decay occurs in the PE spatial encoding direction compared to the frequency-encode (FE) direction due to the longer time difference between adjacent k-space points along the PE direction in comparison to the FE direction. Consequently, T2* blurring will be more significant along the PE direction. For spiral trajectories, the signal decays uniformly in all radial directions. The effective resolution of each protocol was determined in PE direction using the full width at half maximum (FWHM) of the PSFs for the GM and WM.

Two-dimensional PSFs of the simulated k-space data for the WM were calculated on a 4096×4096 grid image using the image reconstruction method described in section 2.1. Shape and magnitude of the main lobe and side lobes affect the contribution of other voxels to the final value of the central voxel, and its contrast with respect to neighbouring voxels. To characterise these effects, we define below the specificity, sharpness, and effective resolution of the PSF. The specificity is defined as the integral of the main lobe within the nominal voxel size in both PE and FE directions normalized by the integral of the rest of the PSF outside the nominal voxel.

$$Specificity = \frac{\sum main\ lobe}{|\sum side\ lobes + residual\ main\ lobe|} \tag{4}$$

While positive side lobes have an overall blurring effect, negative lobes cause sharpening of the resulting image. Sharpness is defined as in (5), of which higher values indicate a greater sharpening effect of the PSF.

$$Sharpness = \frac{|\sum negative\ side\ lobes\ |}{|\sum positive\ side\ lobes + residual\ main\ lobe|} \tag{5}$$

## 2.3 Experiments

### 2.3.1 In-vivo scans to validate simulation results

To validate the simulation results, a volunteer (female, 24 years old) was scanned on a 7T Terra scanner running VE12U-SP01 (Siemens, Erlangen, Germany) using a single channel transmit and 32-channel receive coil (Nova, Wilmington, USA). All scans were approved by the Research Ethics Board of the Montreal Neurological Institute, and informed consent was obtained from all subjects.

Most scan parameters were similar to the Human Connectome Project 7T protocol (Vu et al., 2015). The subject was scanned at a nominal isotropic resolution of 1.5 mm. While only a few (40) slices were acquired to reduce the reconstruction time, the TR of the protocols was set to 5 seconds to avoid signal saturation. All scan parameters are listed in Table 1. Coil sensitivity and $\Delta B_0$ field maps were estimated using a bipolar GRE scan with 6 echos, $TE_1 = 3.81$ ms, and $\Delta TE = 1.07$ ms, in-plane resolution = 1.5 mm covering the same field view as the diffusion scans. A multi-shell diffusion-weighted spin echo protocol was acquired with b-values = 0, 300, 1000, and 2000 s/mm$^2$ in 5, 5, 16, and 48 directions respectively. The TE of all sequences was adjusted for a b-value of 2000 s/mm$^2$.

All scans, including the GRE sequences, were monitored using field monitoring probes in a separate session and the field measurements were used for offline image reconstruction as described in section 2.1 to correct for $B_0$ non-uniformity and gradient imperfections. Motion correction was performed on all images of the partial brain scans with nominal isotropic resolution of 1.5 mm and effective resolution of 1.5 mm using the Multidimensional diffusion MRI (MD-dMRI) (Nilsson et al., 2018) toolbox in MATLAB. No further pre-processing that could impact image resolution (e.g., denoising or Gibbs ringing correction) was performed. Fractional anisotropy (FA) maps were generated from the motion corrected images including all acquired b-values using MRtrix3 (Tournier et al., 2019). We compared the calculated FA maps as opposed to the raw diffusion-weighted images, since differences in the TEs results in differences in the T2-weighted image contrast.

### 2.3.2 In-vivo scans to investigate SNR

Twenty-seven images without diffusion encoding (b-value = 0) were acquired in a volunteer (male, 31 years old) to calculate the SNR of the different readout trajectories at three isotropic nominal resolutions of 1, 1.2, and 1.5 mm with parameters in Table 1. The TR for all scans was matched to the longest TR of the protocols, and the TE was adjusted for a b-value of 2000 s/mm$^2$.

SNR maps were generated using the pseudo multiple replica method (Robson et al., 2008). Briefly, the noise covariance matrix across the receive coil channels was calculated using noise scans added to the onset of the sequences, amounting to 11000 samples in total. One hundred sets of correlated complex-valued Gaussian white noise were generated for each scan with the same dimension as the raw k-space data. To obtain 100 image replicas per scan, the synthesized noise sets were added to the raw k-space data followed by image reconstruction. Images without added noise were also reconstructed to use as *original* scans. A standard deviation (SD) map of the noise for each scan was generated by calculating pixel-wise SD over the stack of replicas. The real part of image replicas was used in calculating the noise SD maps. SNR maps were then estimated as the magnitude of the *original* images divided by the corresponding noise SD map. The final calculated SNR was the average over a WM and GM mask extracted from the b = 0 s/mm$^2$ images.

### 2.3.3 In-vivo scans with matching effective resolution to investigate SNR and efficiency

In order to investigate the SNR and efficiency of different trajectories with a matching effective resolution, whole brain scans of a third volunteer (female, 26 years old) were acquired. The nominal resolution for each scan was chosen using the simulation results for a matching 1.5 mm isotropic effective resolution. TR for every protocol was chosen to minimize the scan time. The other scan parameters are listed in Table 1. In addition to FA maps, intra-cellular volume fraction (ICVF) maps were calculated using neurite orientation dispersion and density imaging (NODDI) (Zhang et al., 2012) to investigate the effect of SNR on microstructural models that require shells with high b-values. Motion corrected diffusion images were denoised using MRTrix3 (Tournier et al., 2019) then ICVF maps were generated using AMICO (Daducci et al., 2015).

**Table 1: In-vivo scan parameters at 7T for three experiments to validate simulation results, calculate SNR, and investigate SNR and efficiency of trajectories with a matching resolution**

| Scan | Nominal resolution | | | SNR measurement | | | | | | | | | Effective resolution | | |
|---|---|---|---|---|---|---|---|---|---|---|---|---|---|---|---|
| Nominal resolution (mm) | 1.5 | 1.5 | 1.5 | 1.5 | | | 1.2 | | | 1 | | | 1.2 | 1 | 1 |
| Trajectory | EPI | PF-EPI | Spiral | EPI | PF-EPI | Spiral | EPI | PF-EPI | Spiral | EPI | PF-EPI | Spiral | EPI | PF-EPI | Spiral |
| TE (ms) | 82,73 | 72,63 | 46 | 102,82,73 | 72,63,59 | 46 | 118,92,80 | 81,67,63 | 46 | 101,87 | 86,72,66 | 46 | 86 | 71 | 46 |
| TR (ms) | 5000 | | | 6700 | | | 6700 | | | 6700 | | | 11100 | 10300 | 8300 |
| FOV (mm³) | 256×256×60 | | | 256×256×36 | | | 256×256×36 | | | 256×256×36 | | | 256×256×144 | | |
| Slice thickness (mm) | 1.5 | | | 1.5 | | | 1.2 | | | 1 | | | 1.5 | | |
| R | 3,4 | 2,3 | 4,5 | 2,3,4 | 2,3,4 | 4,5,6 | 2,3,4 | 2,3,4 | 4,5,6 | 3,4 | 2,3,4 | 4,5,6 | 3 | 4 | 5 |
| PF factor | - | 0.75 | - | - | 0.75 | - | - | 0.75 | - | - | 0.75 | - | - | 0.75 | - |
| Bandwidth-per-pixel (Hz) | 1384 | 1384 | - | 1384 | 1384 | - | 1374 | 1374 | - | 1396 | 1396 | - | 1798 | 1776 | - |
| Number of slices | 40 | | | 24 | | | 30 | | | 36 | | | 96 | | |
| Scan time (min) | ~7 | | | ~1 | | | ~1 | | | ~1 | | | ~15 | ~13 | ~10 |

### 2.3.4 Coil sensitivity and ΔB₀ field map estimation, and image reconstruction

Individual coil images from the GRE scan were first reconstructed by explicit multiplication of the Hermitian conjugate of the encoding matrix in (2) excluding the $\Delta B_0$ term. Coil sensitivity maps were estimated from the first echo using ESPiRIT (Uecker et al., 2014). To map the $B_0$ non-uniformity, pixel-wise unwrapping of the phase image of each channel across all echoes was performed, followed by averaging $\Delta B_0$ maps obtained for every coil and smoothing the final map using a 7×7×7-pixel spatial median filter.

Measured trajectories up to the 3rd order of spherical harmonics, coil sensitivity maps, and $\Delta B_0$ maps were used in the expanded signal model in (1) to reconstruct the diffusion-weighted images.

### 2.3.5 Eddy current compensation

The Siemens scanner data acquisition pipeline includes online eddy current compensation (ECC) that adjusts the system's central frequency $f_0$ during the signal demodulation, which adds a phase term to the raw data. This correction needs to be disabled since these $f_0$ variations are also measured by the field probes, otherwise eddy current effects will be corrected twice during image reconstruction. Since this feature cannot be disabled on the 7T Terra scanner, we must invert the scanner's ECC. The same protocols were simulated in the IDEA environment to obtain gradient waveforms, which were converted to ISMRMRD format[1] to calculate the ECC applied by the scanner to the raw data in the form of a $k_0$ phase terms. The scanner ECC correction is inverted by multiplying the raw data by the conjugate values of ECC phase terms. The measured $k_0$ terms obtained by field probe measurements, which are more accurate than scanner's simulated eddy currents, are applied instead during the image reconstruction.

## 3 Results

### 3.1 Simulation results

#### 3.1.1 Sequence timing

Figure 1A and B show the readout duration and echo time of the simulated trajectories as a function of nominal resolution, respectively. The readout duration of the spiral trajectory with R

---

[1] https://github.com/SkopeMagneticResonanceTechnologies/siemens_to_ismrmrd

= 4 is shorter compared to EPI with the same acceleration factor for almost all resolutions, and it is shorter than PF-EPI for resolutions lower than 1 mm. The rate at which the readout duration increases at high-resolutions is greater for spirals than for PF-EPI and EPI due to the radial pattern of k-space acquisition in spiral trajectories.

Echo times in Figure 1B were calculated for sequences with a b-value of 2000 s/mm². In spiral trajectories, the TE is independent from the resolution and remains at 44 ms over the entire range. The echo time of EPI and PF-EPI increases with resolution as expected. Results show a significant advantage of spiral trajectories over EPI-based trajectories due to the shorter TE resulting in a higher SNR, particularly at high resolutions.

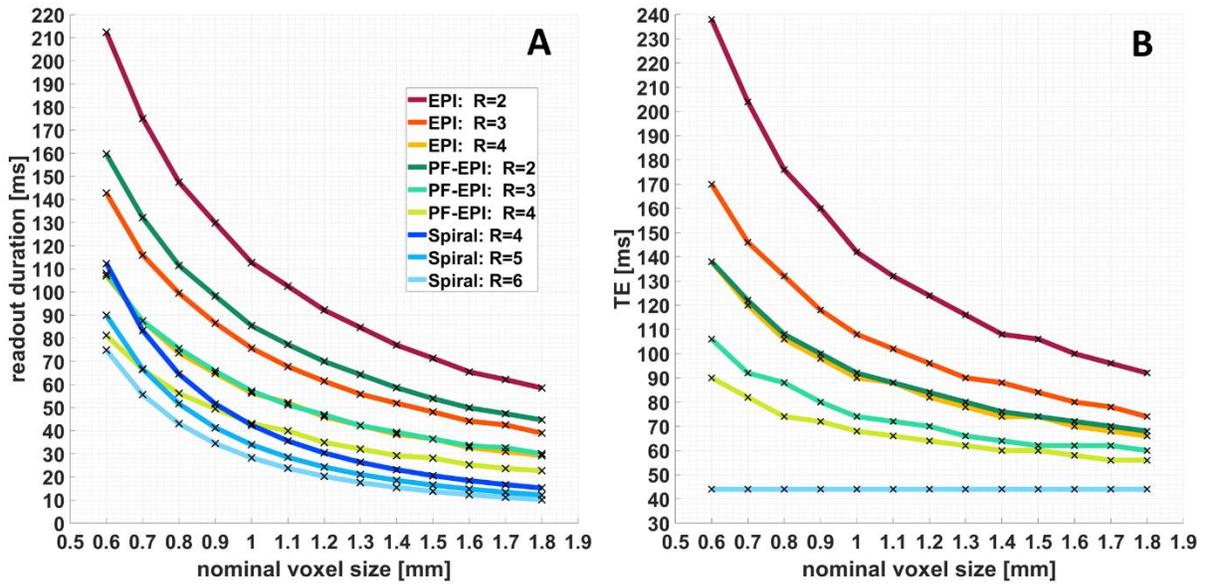

***Figure 1- Timing properties of trajectories.*** *A: Readout duration as a function of nominal resolution. Spiral trajectories have shorter readout durations due to more efficient way of sampling k-space for the same acceleration factor. B: Echo time (TE) as a function of nominal resolution for a b-value = 2000 s/mm². The echo time remains at 44 ms for spiral trajectories, while it increases with resolution for EPI and PF-EPI.*

### 3.1.2 Point spread function

The modulation transfer function (MTF) along the PE axis, reflecting the T2 and T2* signal decay along the readout trajectory, and the corresponding PSFs for EPI, PF-EPI, and spiral trajectories are shown in Figure 2 for WM and GM. Equivalent simulation results at 3T are included in Figure S1 of supplementary material for comparison. MTF signal amplitude was normalized so that the value at $k_y = 0$ is one. The right column of Figure 2 shows one-sided PSFs calculated from Fourier transformation of the corresponding MTFs. There are large variations in PSFs between the readout trajectories. Due to the shorter T2* time of the WM in comparison to the GM, PSFs are wider for the WM. This broadening of PSFs indicates more blurring, which results in a lower effective resolution. On the other hand, as the PSF gets sharper, the amplitude of associated side lobes becomes larger, which affects specificity and sharpness.

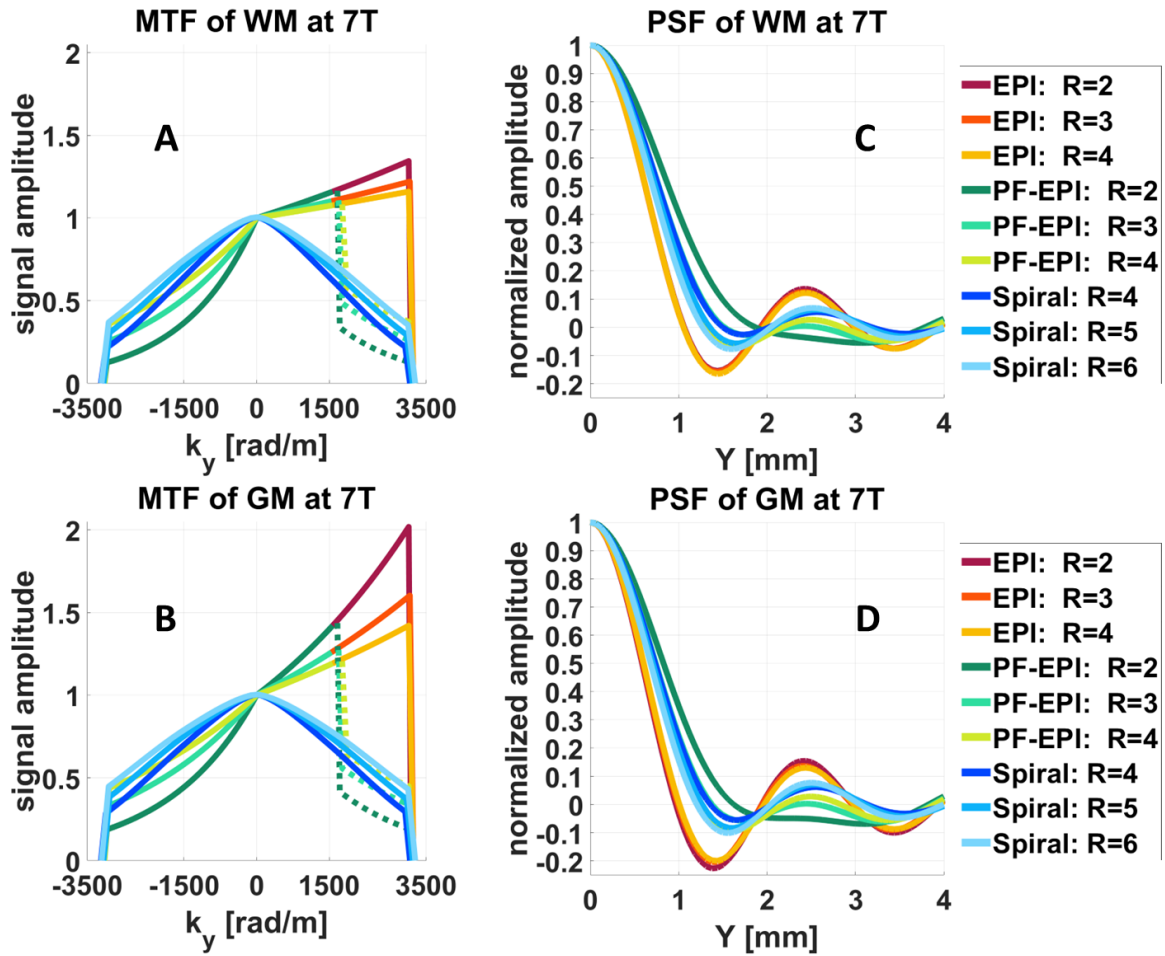

***Figure 2- MTF along the PE direction and corresponding PSF.*** *A and B, and their corresponding PSFs in C and D for the WM and GM at 7T. There is more broadening of the PSF for the WM in comparison to the GM. The dashed portion of the PF-EPI MTFs was generated using the Hermitian conjugate property of the k-space.*

The real part of the 2-D PSFs of the WM for a nominal resolution of 1 mm at 7T are shown in Figure 3A and B for spiral and EPI trajectories, respectively. For EPI, ringing amplitudes are greater along the PE and FE axes while spiral has circular ringing that uniformly spreads in all radial directions. The effective resolution is compared to the nominal resolution at 7T in Figure 3C and D, and at 3T in Figure S2 of the supplementary material. There are large variations in the effective resolution between the trajectories at 7T. As expected, the WM has a lower effective resolution than the GM due to its shorter T2* time. EPI and PF-EPI trajectories follow a linear trend over the range of resolutions considered, while spiral trajectories show a deviation from linearity for resolutions higher than 0.9 mm due to the extensive signal loss caused by long readout durations. This results in the suppression of high frequency components and thus a lower effective resolution. The specificity for all 2-D PSFs, defined in section 2.2, are shown in Figure 3E. The specificity decreases at higher resolutions for all protocols. EPI has the highest specificity, which is expected due to its sharper peak, as shown in 1-D PSFs in Figure 2. The decrease in specificity at higher resolutions is most significant for spiral trajectories due to excessive suppression of high frequencies by the T2* signal decay. The sharpening effects of EPI and PF-EPI remain almost constant over different resolutions, while this effect is significantly reduced at high resolutions for spirals as shown in Figure 3F. This is due to the signal decay which causes suppression of higher frequencies leading to decreasing side lobe amplitudes, while the residual main lobe remains at a high positive value. This sharpening effect in EPI and PF-EPI causes Gibbs ringing artifacts in the image, while spirals inherently reduce them, specifically at high resolutions.

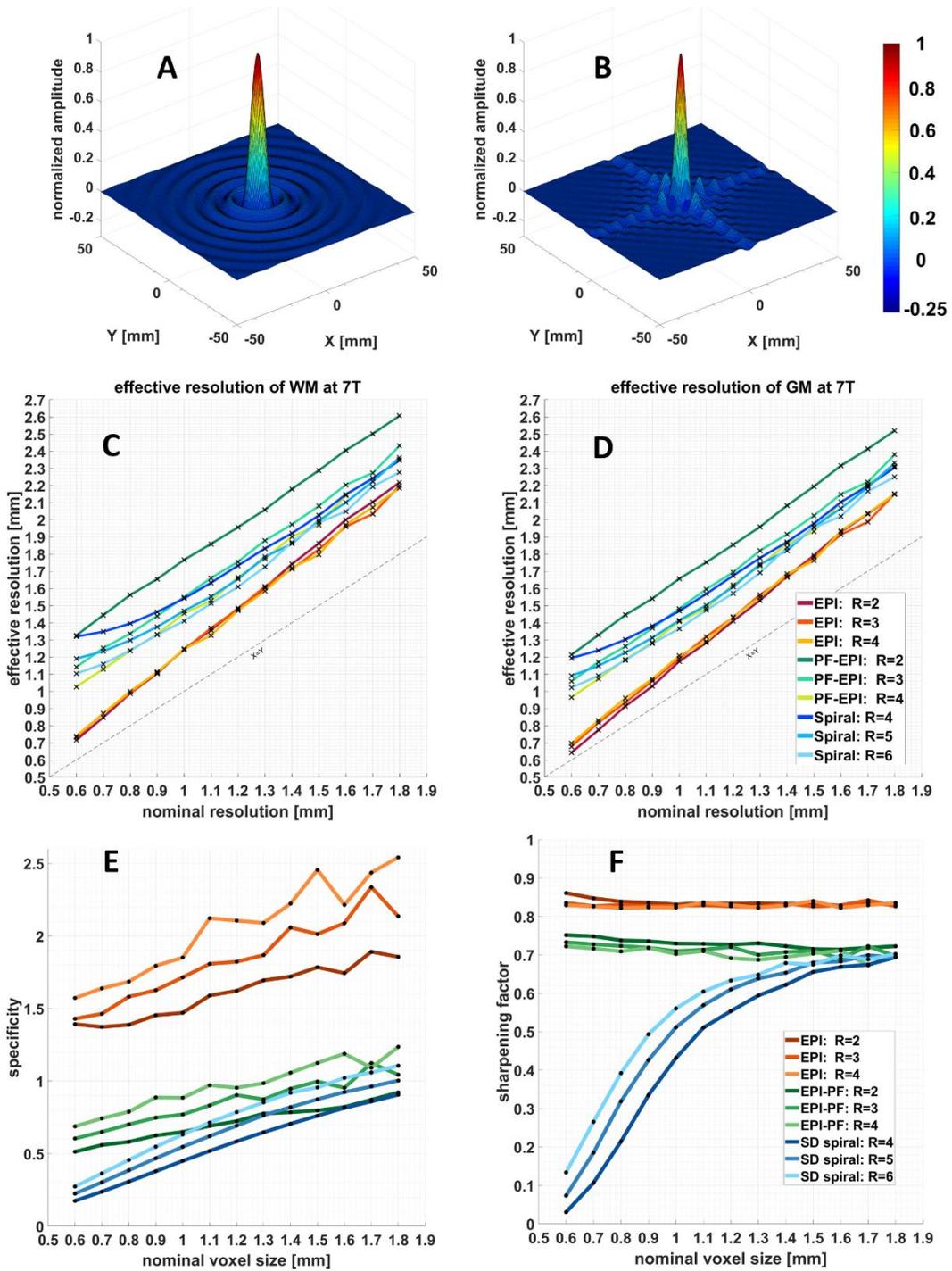

***Figure 3: PSF analysis.*** *Spiral has similar ringing in all directions while ringing is constrained along the PE and FE axes for EPI. The specificity of EPI is higher due to its narrower main lobe compared to PF-EPI and spiral. EPI and PF-EPI have a constant sharpening effect, while the sharpness of spirals reduces significantly at high resolutions due to the signal decay causing suppression of the side lobes.*

In addition to PSF simulations, a digital brain phantom was simulated to study effects of CG image reconstruction on image quality. Methods and results can be found in supplementary material. The results are similar to the PSF analysis results described above.

## 3.2 In-vivo scan results

### 3.2.1 EPI has the highest effective resolution

Figure 4 shows FA maps derived from the 1.5-mm scans shown in Figure 3S of supplementary material in the axial, sagittal, and coronal planes. The SNR advantage of spirals over EPI-based trajectories is clearly visible in the mean DWI images. The color FA maps of the PF-EPI scans clearly show blurring of fine structures in comparison to EPI and spirals along the anterior-posterior direction. In contrast with spiral trajectories characterized by uniform blurring in all directions in-plane, the majority of blurring due to T2* decay in EPI-based trajectories appears along the phase-encoding (PE) direction, here the anterior-posterior direction. It is therefore expected to see a maximal blurring in the sagittal and axial planes, and minimal blurring in the coronal plane. A clear example of this in the axial and sagittal planes is the corticospinal fibers that form a striping pattern in the PE direction and are affected the most by the blurring.

EPI trajectories provide the sharpest FA maps in the sagittal and axial planes, in particular for R = 4 due to the shorter readout and thus less T2* decay. The spiral with R = 5 shows slightly sharper FA maps compared to the spiral with R = 4 and PF-EPI with R = 3. The blurriest FA map is obtained by using PF-EPI scans with R = 2, mainly due to its longer readout duration.

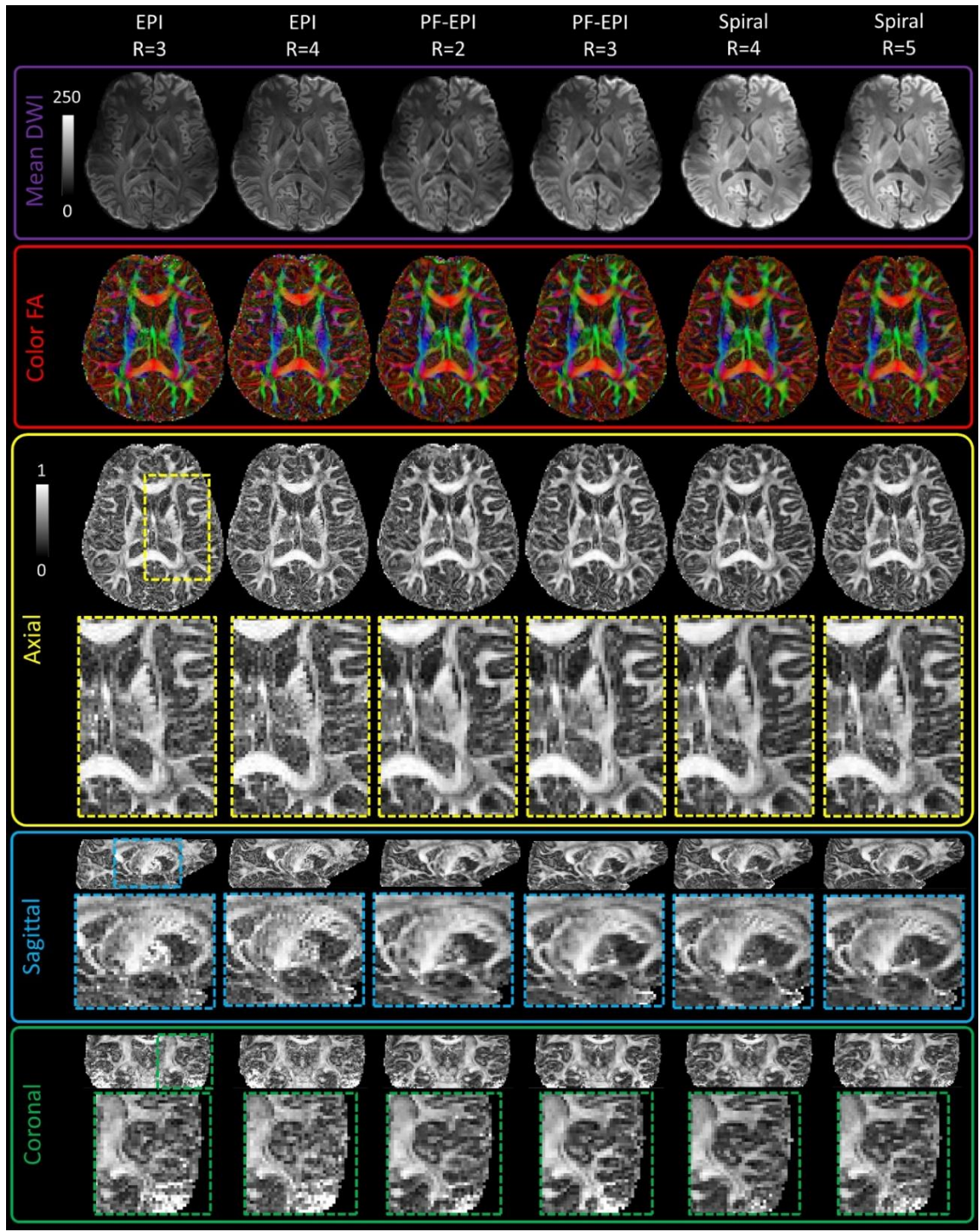

***Figure 4- FA maps calculated using different trajectories at 1.5 mm isotropic nominal resolution.*** *Color FA maps in the axial plane are shown in the first row, and FA maps in axial, sagittal and coronal planes with magnified regions for better examination are shown below. EPI-based scans show a minimal blurring in the coronal direction, and maximal blurring in the sagittal plane, while blurring in the spiral trajectory occurs in all directions. The sharpest FA map is acquired using EPI with R = 4, and the map with the lowest effective resolution is generated using PF-EPI with R = 2.*

To investigate the blurring effects on the calculated maps, FA values and smoothness of structures in specific regions of interest selected along the FE (Figure 5A, B) and PE (Figure 5C, D) axes in the 1.5-mm isotropic and 1-mm anisotropic scans were investigated more closely using line plots. These regions were selected to include fibers oriented perpendicular to the ROI. In Figure 5A and B, FA values obtained using EPI and PF-EPI trajectories are consistent within a range of ~0.1. Spirals show smoother FA profiles and larger differences compared to EPI and PF-EPI, as highlighted by the blue arrow in Figure 5A. In Figure 5C and D corresponding to the PE direction, the difference in FA values between EPI and PF-EPI trajectories is more significant than in the FE direction. These plots show sharper changes in FA for EPI trajectories and smoother variations for PF-EPI and spiral trajectories.

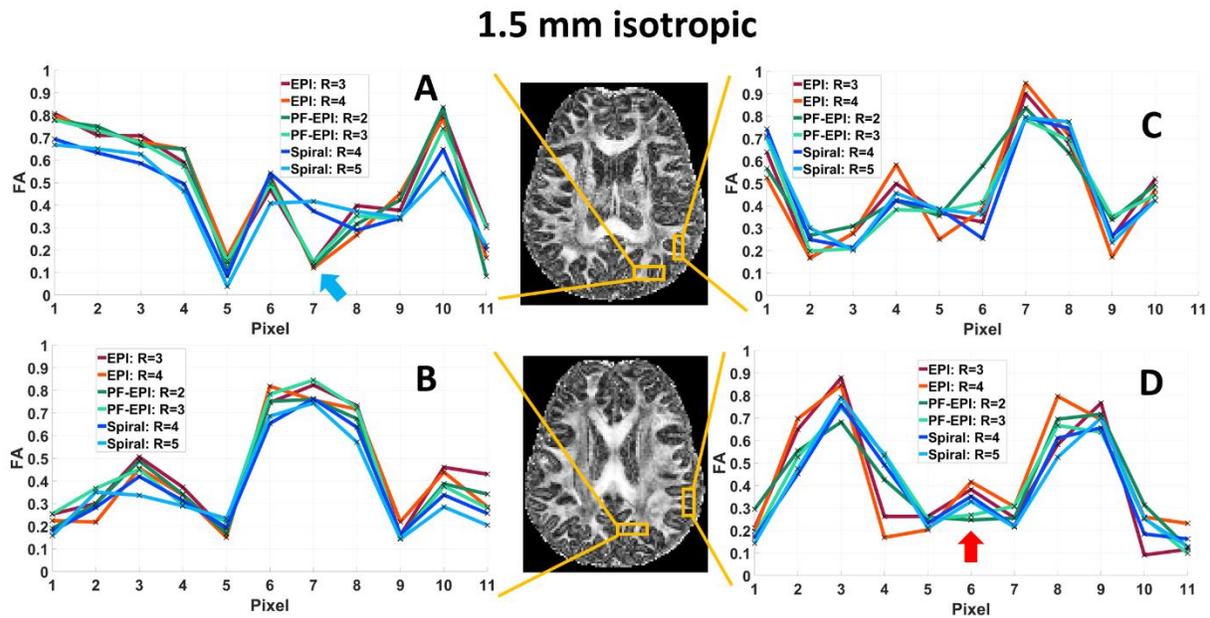

**Figure 5- Line plots of FA values in PE and FE directions at a nominal isotropic resolution of 1.5 mm.** *A and B: line plots of FA values along the FE direction. C and D: line plots of FA values along PE direction. FA values show more variations in EPI and PF-EPI trajectories in the PE direction compared to the FE direction. The blue arrow shows more variability of the FA values calculated using spirals in the FE direction. The red arrow shows a drop in FA for PF-EPI with R = 2 and 3 in contrast to other trajectories.*

### 3.2.2 Spirals provide the highest SNR

The SNR values calculated from the in-vivo scans using different trajectories and parameters at three isotropic resolutions of 1, 1.2, and 1.5 mm are plotted in Figure 6. EPI with R = 2 at 1 mm was excluded due to low signal amplitude of the field monitoring probes towards the end of the readout. For a given acceleration factor R, EPI has the lowest SNR, mainly due to its longer echo time. The SNR of spiral trajectories varies the most as a function of R due to changes in the under-sampling rate in two dimensions compared to EPI and PF-EPI. Furthermore, the echo time of EPI and PF-EPI is shortened at higher acceleration factors which partially compensates for the SNR loss due to the increased undersampling. This figure clearly shows the advantage of spirals in preserving a high SNR at high resolutions.

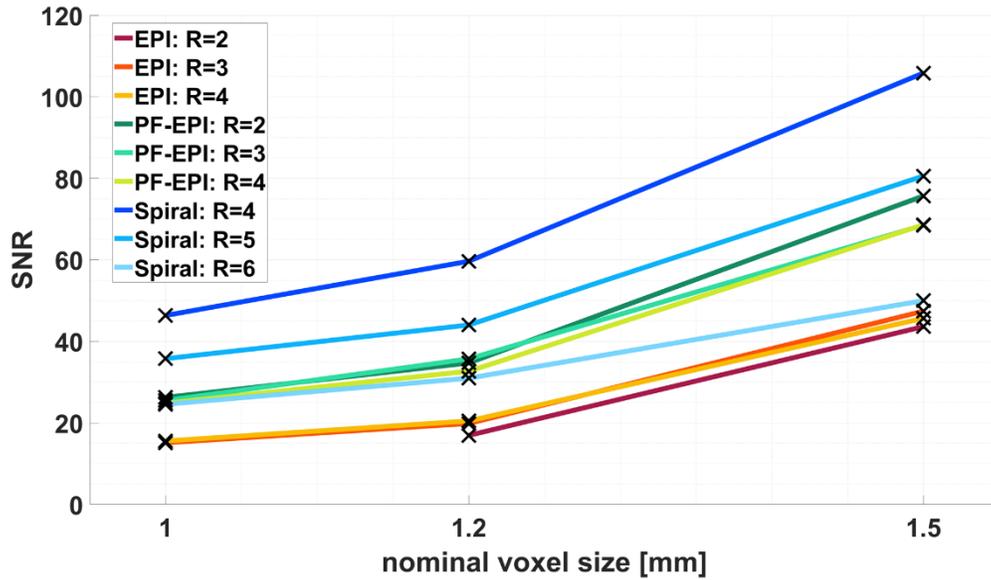

***Figure 6- SNR calculated from in-vivo scans.*** *The SNR was calculated within a brain tissue mask in b=0 s/mm² scans at three resolutions of 1, 1.2, and 1.5 mm isotropic.*

### 3.2.3 Spirals provide highest SNR for matching effective resolution

FA and ICVF maps calculated from whole brain scans (Figure S4 of supplementary material) with a matching effective resolution of 1.5 mm are shown in Figure 7. The SNR of the b=0 s/mm² images for EPI, PF-EPI, and spiral were 23.7, 18.8, and 32.4, respectively. Despite the higher nominal resolution of the spiral trajectory to match the effective resolution of the other images, the SNR of spirals is still higher than for EPI. Although all scans provide FA maps of similar quality, ICVF maps clearly show the advantage of the higher SNR of the spirals for the shells with high b-values of 2000 s/mm². Furthermore, the spirals shorten the scan time by about 33% and 23% compared to EPI and PF-EPI, respectively. FA and ICVF maps of different slices are available in Figures S5 and S6 of supplementary material, respectively.

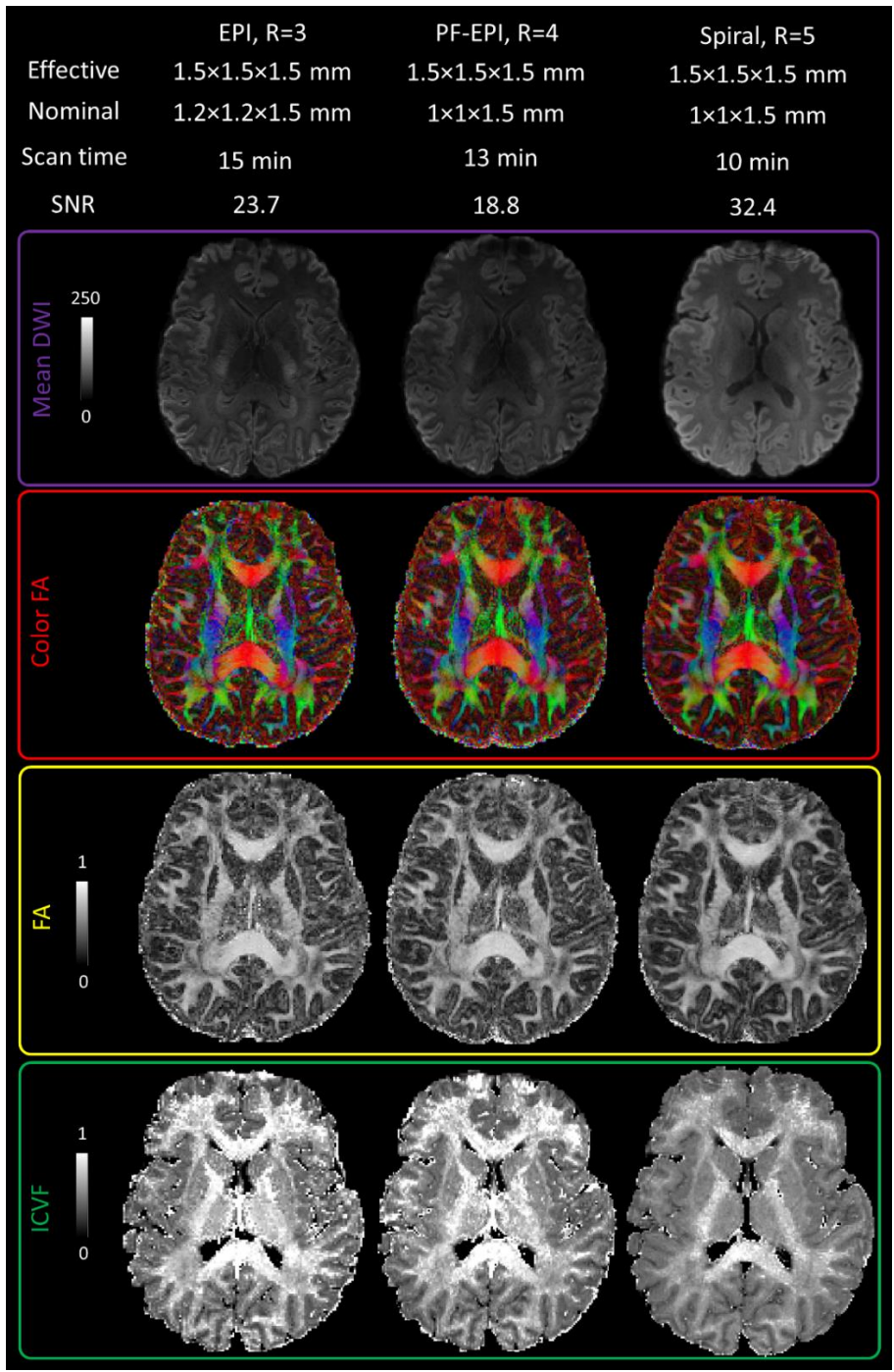

**Figure 7- FA map of scans with a matching effective resolution of 1.5 mm.** *Similar structures in FA maps can be seen in all maps due to the matching effective resolution. Effect of higher SNR of the spiral scan is clear in ICVF maps.*

**4 Discussion4.1 Spirals are the optimal k-space readout trajectory for single-shot dMRI at 7T**

The aim of this study was to characterise the effects of T2* decay on spatial resolution and quality of dMRI at 7T and to find an optimal single-shot readout trajectory that balances the trade-off between SNR and image resolution. We characterised the PSF of dMRI with EPI, PF-EPI, and spiral trajectories using sequence simulations. Three measures were proposed for comparison of the trajectory PSFs: specificity, sharpness, and effective resolution. In vivo scans were acquired at 7T to investigate consistency with simulation results, as well as to measure SNR. Field monitoring probes were used to eliminate distortions and artifacts caused by field imperfections. We showed that spirals generally have lower effective resolution and specificity compared to EPI at matching nominal resolutions. However, the SNR advantage of spiral enables the acquisition of single-shot spiral dMRI scans at an effective resolution of ~1.5 mm for a b-value of 2000 s/mm$^2$ at a higher SNR and in a shorter scan time than EPI and PF-EPI.

**4.2 Spatial specificity and sharpening factor**

The PSF is typically characterised using the FWHM. Engel et el. (2022) have recently used specificity and sensitivity in addition to FWHM to characterize the PSF and determine the optimal TE for BOLD fMRI contrast using spiral and EPI trajectories. They defined specificity as the ratio between the integral over the main lobe and the L$^2$-norm of the side lobes. Here we used a different definition for specificity: the ratio between the main lobe within the nominal voxel boundaries to the integral of the side lobes. This definition was used to better reflect the contribution of spins within the nominal voxel. This specificity measure is affected by the residual main lobe, where a sharper peak in the PSF leads to a reduction of the area under the residual main lobe. This is the main reason that EPI has the greatest specificity, even though its side lobes have higher amplitude than the other trajectories. Spirals have more variable specificity over the range of resolutions studied; higher suppression of side lobes leads to lowering side lobe amplitudes significantly. Side lobe suppression is expected to increase specificity; however, the greater area of the residual main lobe of spirals dominates and reduces the specificity.

The sharpness quantifies the effects of negative side lobes on image quality. A greater sharpening effect is not necessarily advantageous since it increases Gibbs ringing and intensifies edges. Due to suppression of high-frequency components using spirals, it has an inherent benefit of removing Gibbs ringing, especially at high resolutions.

**4.3 Simulation results of the effective resolution are consistent with in-vivo scans**

Simulation results in Figure 3 clearly show differences in the effective resolution between different trajectories, which are enhanced at higher resolutions. At 1.5-mm nominal resolution, the effective resolution can be nearly divided into three different groups where EPI trajectories perform best, and PF-EPI with R=2 have the lowest effective resolution, and other trajectories in between. In vivo FA maps in Figure 4 and line plots of Figure 5 confirm these considerable differences observed in the simulations, more specifically perpendicular to structures oriented along the FE direction, such as the corticospinal tract.

**4.4 Trade-off between SNR and effective resolution**

Several groups have investigated the gain in SNR at higher field strengths for diffusion MRI (Choi et al., 2011; Reischauer et al., 2012). In a recent study at 3T, Lee et al. (2021b) used field monitoring probes and measured the SNR benefit of spiral over EPI trajectories. They performed a PSF analysis for trajectories with an equivalent effective resolution of 1 mm. Reported SNR values are lower than what we calculated in this study by a factor of ~6 for b=0 s/mm² at similar TE values. This SNR difference is due to imaging at 7T which is expected to provide ~5.21 ($SNR \propto \sim B_0^{1.95}$) times higher SNR than 3T according to (Pohmann et al., 2016).

As mentioned above, T2 and T2* are approximately halved at 7T compared to 3T. Although we did not perform in-vivo experiments at 3T to compare them to our 7T results, simulations shown in Figure 3 and Figure S2 of supplementary material suggest increased blurring at 7T and greater differences between the different trajectories. Effective resolution of PF-EPI and spiral are decreased ~20% compared to 3T, and differences in the effective resolution among trajectories was increased from ~13% at 3T to ~27% at 7T. Given the greater effect of T2* blurring at 7T, nominal resolutions presented in dMRI studies at 7T should be interpreted with caution, in particular for studies that investigate fine structures of the brain such as the cortical gray matter.

Future work could focus on minimizing the effect of T2* blurring by demodulating the k-space data before image reconstruction using a T2* map, at the cost of enhancing high-frequency noise. The PSF analysis can also be integrated into trajectory optimization methods to find a readout trajectory that minimizes blurring while preserving the SNR (e.g., Weiss et al., 2021).

## 4.5 Diffusion-encoding effects

Different diffusion-encoding strengths (b-values) and schemes (linear, b-tensor) affect TE and therefore potentially the effective image resolution and the SNR. We calculated the effective resolution for various echo times in the PSF analysis and obtained the same results as shown in Figure S6 of the supplementary material. T2* decay after the echo in a spin-echo sequence remains the same regardless of the echo time. However, in a gradient-echo sequence, changes in TE affect the T2* decay modulation and therefore the effective resolution of the scans (Engel et al., 2018)

As shown in Figure S7 of the supplementary material, the differences in echo time for b-values of 500 and 1000 s/mm$^s$ is very small for EPI and PF-EPI readouts compared to spirals. This is due to the added idle time in EPI-based dMRI sequences which in addition to diffusion gradient duration, affects calculation of the b-value, whereas in spirals there is no idle time between diffusion gradients and the refocusing pulse, therefore changes in the b-value depend only on the diffusion gradient duration.

## 4.6 Limitations

The PSF analysis and digital phantom simulations aim to quantify the effects of the different trajectories on the effective resolution and spatial accuracy. An aspect that is not taken into account in the simulations is the variability in tissue properties, such as proton density and relaxation times, across the brain. Despite these limitations, the simulation results at both high- and low- resolutions are consistent with the in-vivo scan results.

There are several techniques available to reconstruct PF-EPI scans such as projection onto convex sets (POCS) (Haacke et al., 1991) and the *virtual coil* concept (Blaimer et al., 2009). The default PF reconstruction mode on Siemens scanners is zero-filling k-space. In this study, we included only the acquired part of k-space in the image reconstruction of PF-EPI scans, which results in similar quality to k-space zero-filling. Other techniques can be used to improve the quality and decrease blurring in PF-EPI, as discussed in previous studies (Huber et al., 2017). Also, more advanced image reconstruction techniques such as LORAKS (Haldar, 2014) can be employed for higher PF factors which reduces the blurring significantly, but at the cost of increased reconstruction time.

To minimize image artifacts and blurring due to field imperfections in the in-vivo scans, we used off-line field measurements. Motion and breathing can cause changes in the zeroth order fields. These effects are negligible for single-shot imaging due to the short readout duration for each slice (~100 ms). However, subject motion could lead to changes in the static $\Delta B_0$ map that is used for image reconstruction. To minimize the discrepancy between the $\Delta B_0$ map used for the image reconstruction and actual $B_0$ non-uniformity, GRE scans were repeated about every 15 minutes.

SMS acquisition was not implemented in this study, which increased the acquisition time. When using SMS with spirals, the scan time is reduced significantly which makes high spatial resolution and high angular resolution whole brain imaging feasible.**5. Conclusion**
The effective resolution achieved using a specific k-space trajectory should be considered as it is significantly lower than the nominal resolution entered at the scanner and typically reported in the literature, in particular at UHFs due to the shorter T2* times of brain tissue. If time is not a limiting factor, multi-shot acquisitions may be preferable as they provide higher SNR, better effective resolution, and specificity.  In this work, we investigated fast, single-shot protocols that can be used in clinical research. We showed that spirals provide sufficient signal to achieve higher effective resolutions than EPI overall and within a shorter scan time.

**Code and data availability**

The MATLAB code used for sequence simulation are available at [https://github.com/TardifLab/dMRI_sequence_simulations]. The image reconstruction pipeline described in section 2.1 is available at [https://github.com/TardifLab/ESM_image_reconstruction]. Raw reconstructed diffusion images and calculated maps are available at [https://doi.org/10.5683/SP3/V7ITEH].

**Acknowledgments**

The authors would like to thank Ronaldo Lopez and David Costa at the McConnell Brain Imaging Centre for helping with the human scans, and Marcus Couch (Siemens Collaboration Scientist) for his technical support. We would like to thank Cameron Cushing and Paul Weavers (Skope MR Inc, WI, USA), and Christian Mirkes (Skope Magnetic Resonance Technologies AG, Zurich) for their technical support for the field monitoring probes.

Funding: This project was funded by the Natural Sciences and Engineering Research Council of Canada, the Fonds de recherche du Québec – Santé, and Healthy Brains for Healthy Lives. The data was acquired at the McConnell Brain Imaging Centre, which is supported by the Canadian Foundation for Innovation, Brain Canada, and Healthy Brains for Health Lives.

# High-resolution diffusion-weighted imaging at 7 Tesla: single-shot readout trajectories and their impact on signal-to-noise ratio, spatial resolution and accuracy

## Supplementary material

Sajjad Feizollah* [a, b], Christine L. Tardif [a, b, c]

a. Department of Neurology and Neurosurgery, Faculty of Medicine and Health Sciences, McGill University, 3801 Rue University, Montreal, QC, Canada

b. McConnell Brain Imaging Centre, Montreal Neurological Institute, McGill University, 3801 Rue University, Montreal, QC, Canada

c. Department of Biomedical Engineering, Faculty of Medicine and Health Sciences, McGill University, Duff Medical Building, 3775 Rue University, Suite 316, Montreal, QC, Canada

* Corresponding author: Sajjad Feizollah: sajjad.feizollah@mail.mcgill.ca

Christine Lucas Tardif: christine.tardif@mcgill.ca

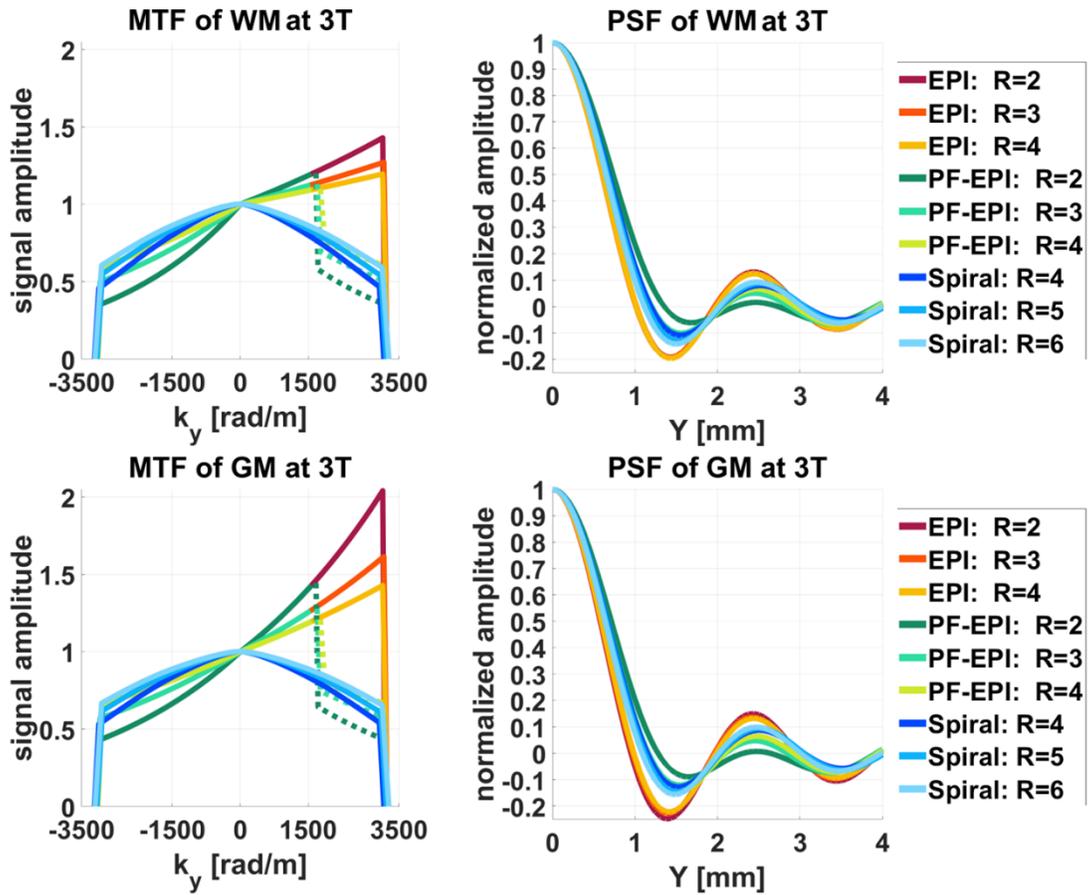

***Figure S8- MTF along the PE direction and corresponding PSF.*** *MTF in the left column, and their corresponding PSFs in right column for the WM and GM at 3T. There is more broadening of the PSF for the WM in comparison to the GM. The dashed portion of the PF-EPI MTFs was generated using the Hermitian conjugate property of the k-space.*

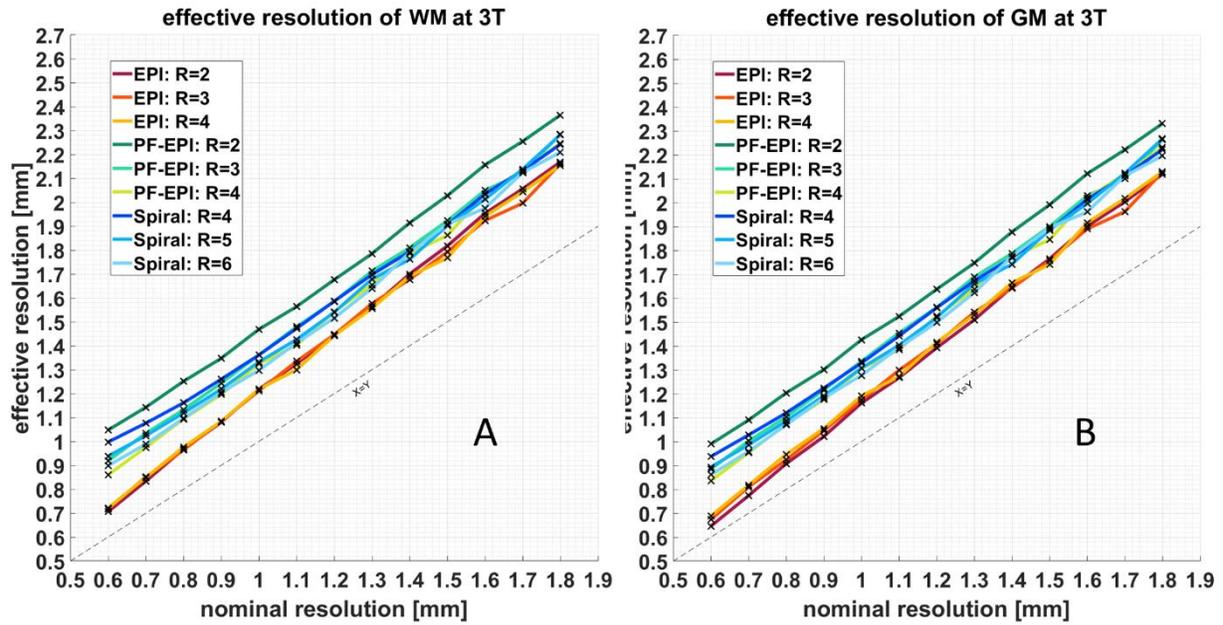

***Figure S9- Effective resolution as a function of nominal resolution.*** *The effective resolution of WM and GM is shown in A and B. WM has a lower effective resolution due to its shorter T2\* relaxation time.*

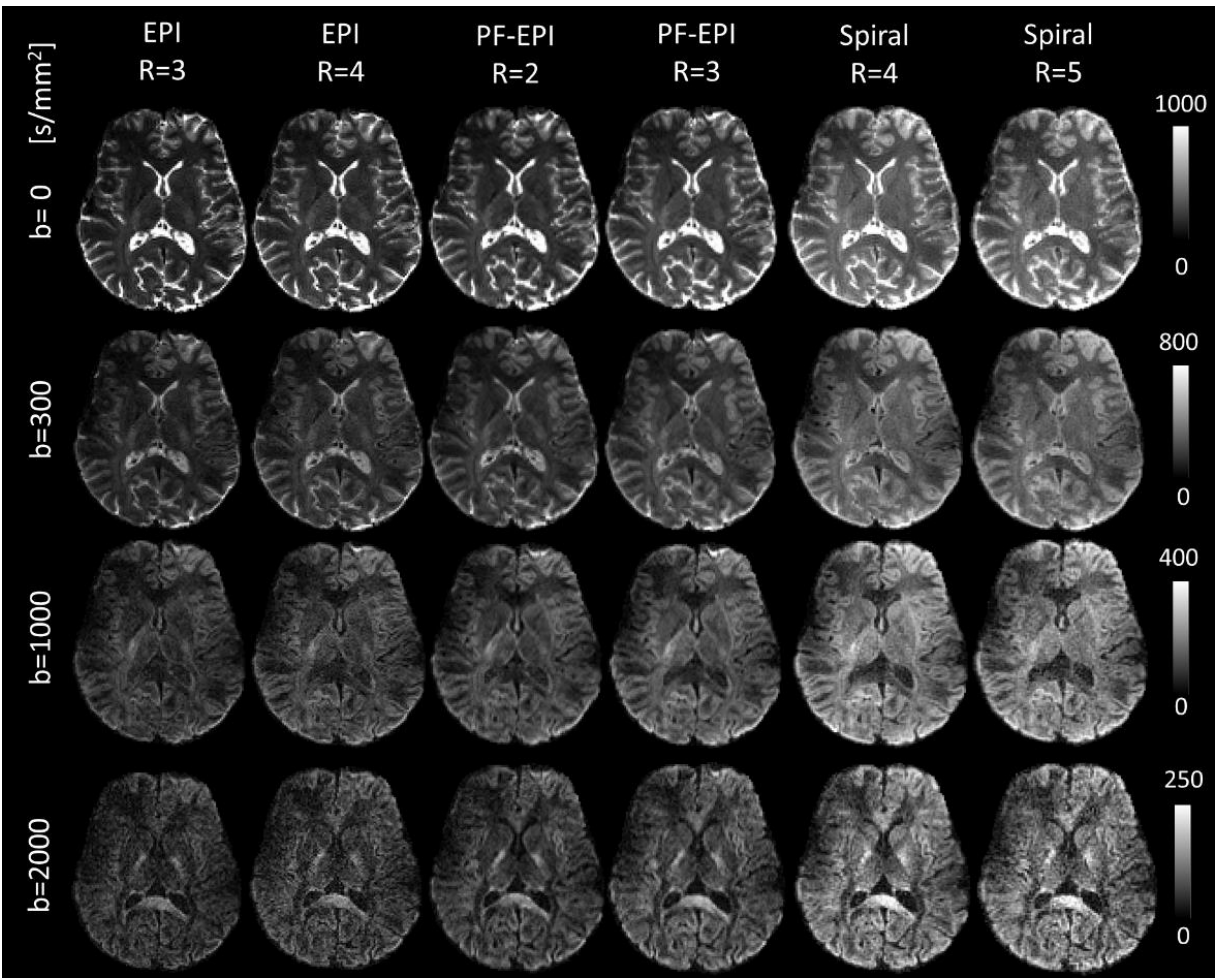

***Figure S10- Diffusion-weighted images using different readout trajectories at a nominal resolution of 1.5 mm isotropic at 7T.*** *Images for b-values of 0, 300, 1000, and 2000 s/mm²  acquired by different trajectories are shown. The mean DWI images in the last row are calculated using 64 directions across shells. Images in the same rows are shown with the same scale adjusted for better visibility.*

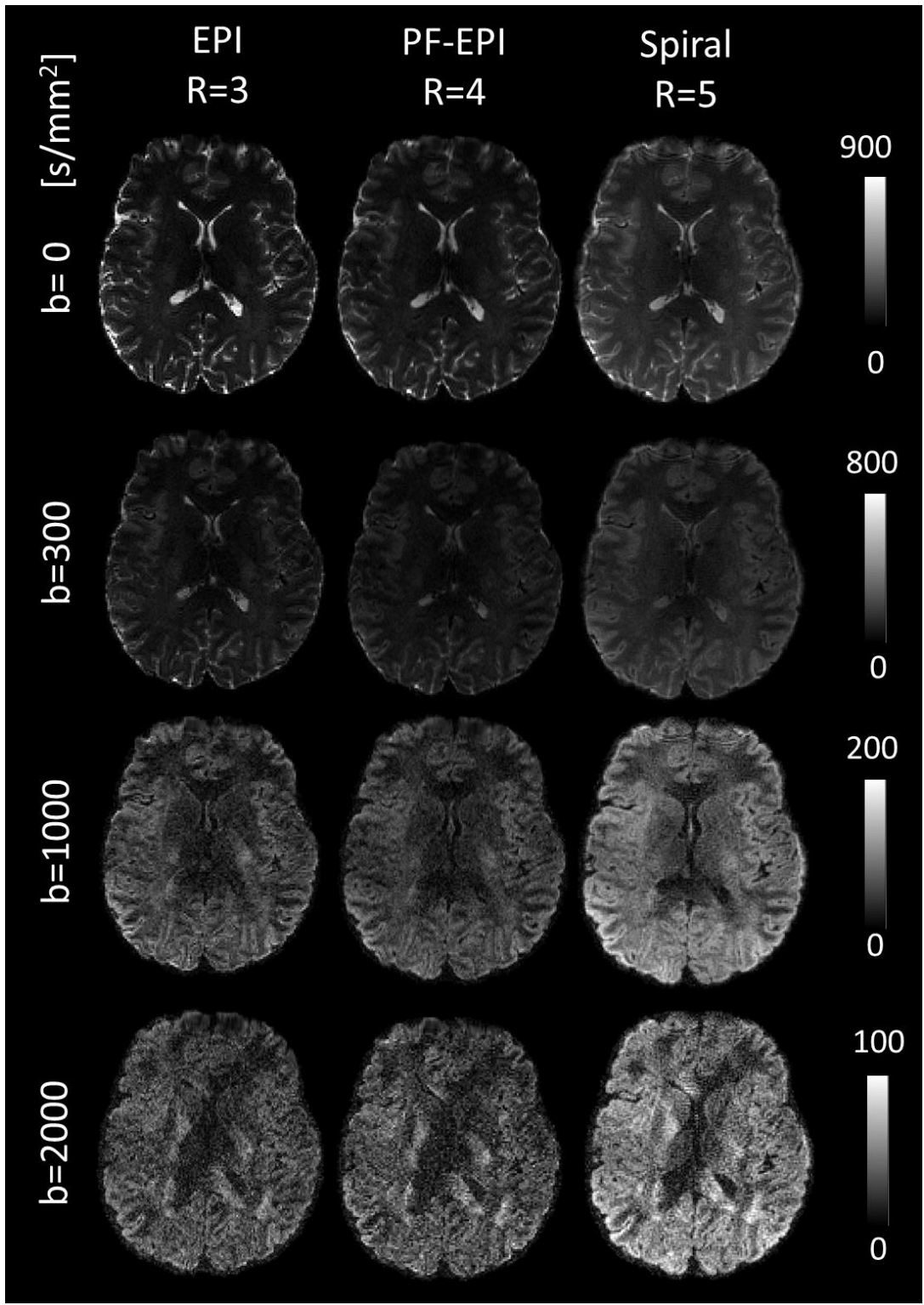

**Figure S11- Reconstructed images at matching effective resolution of 1.5-mm.** *Diffusion images with b-values of 0, 300, 1000, and 2000 s/mm² using different trajectories are shown.*

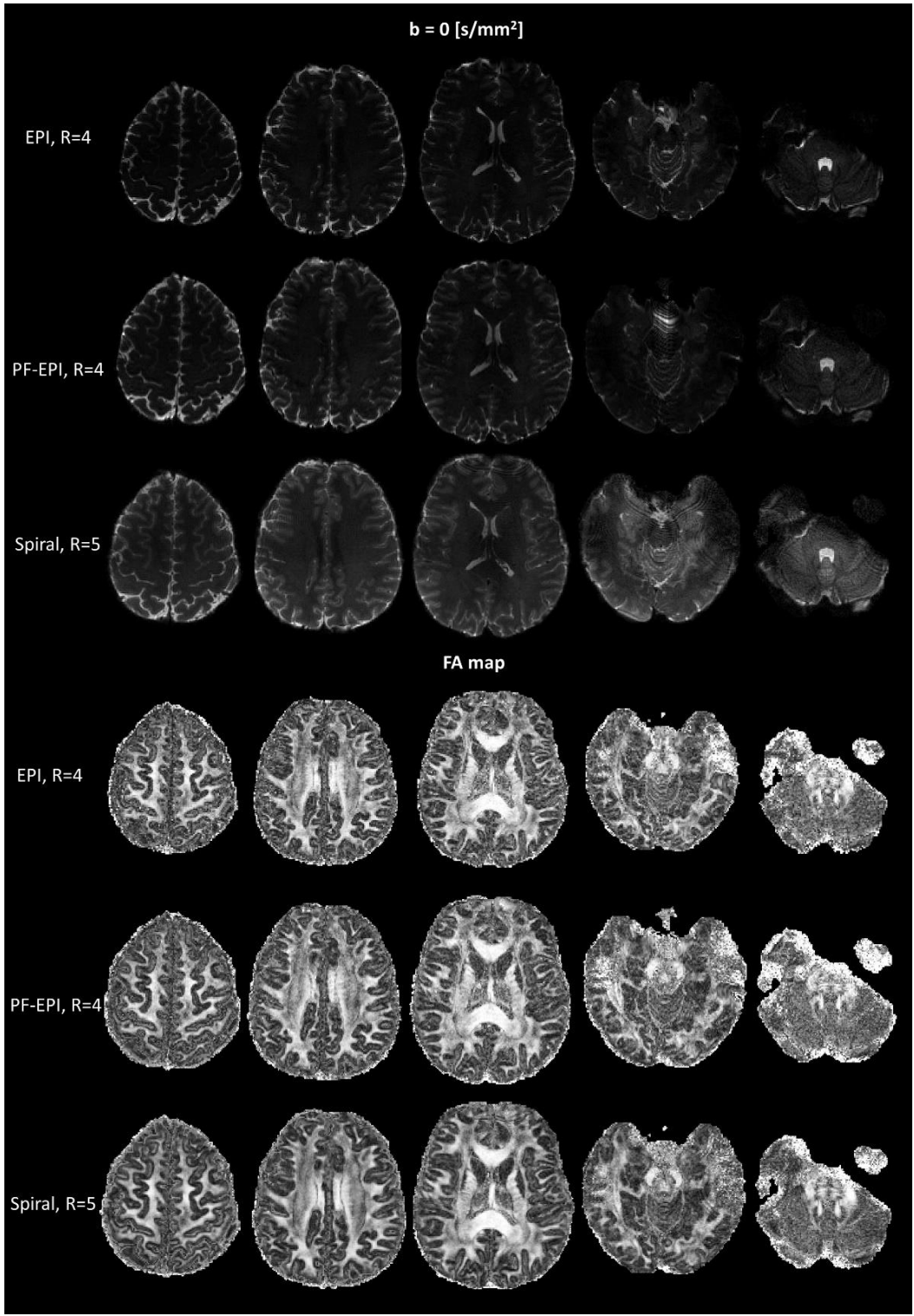

*Figure S12- different slices of b=0 s/mm² and FA maps at matching resolution of 1.5 mm. Effects of $B_1$ nonuniformity are clear in the last two columns which causes loss of SNR.*

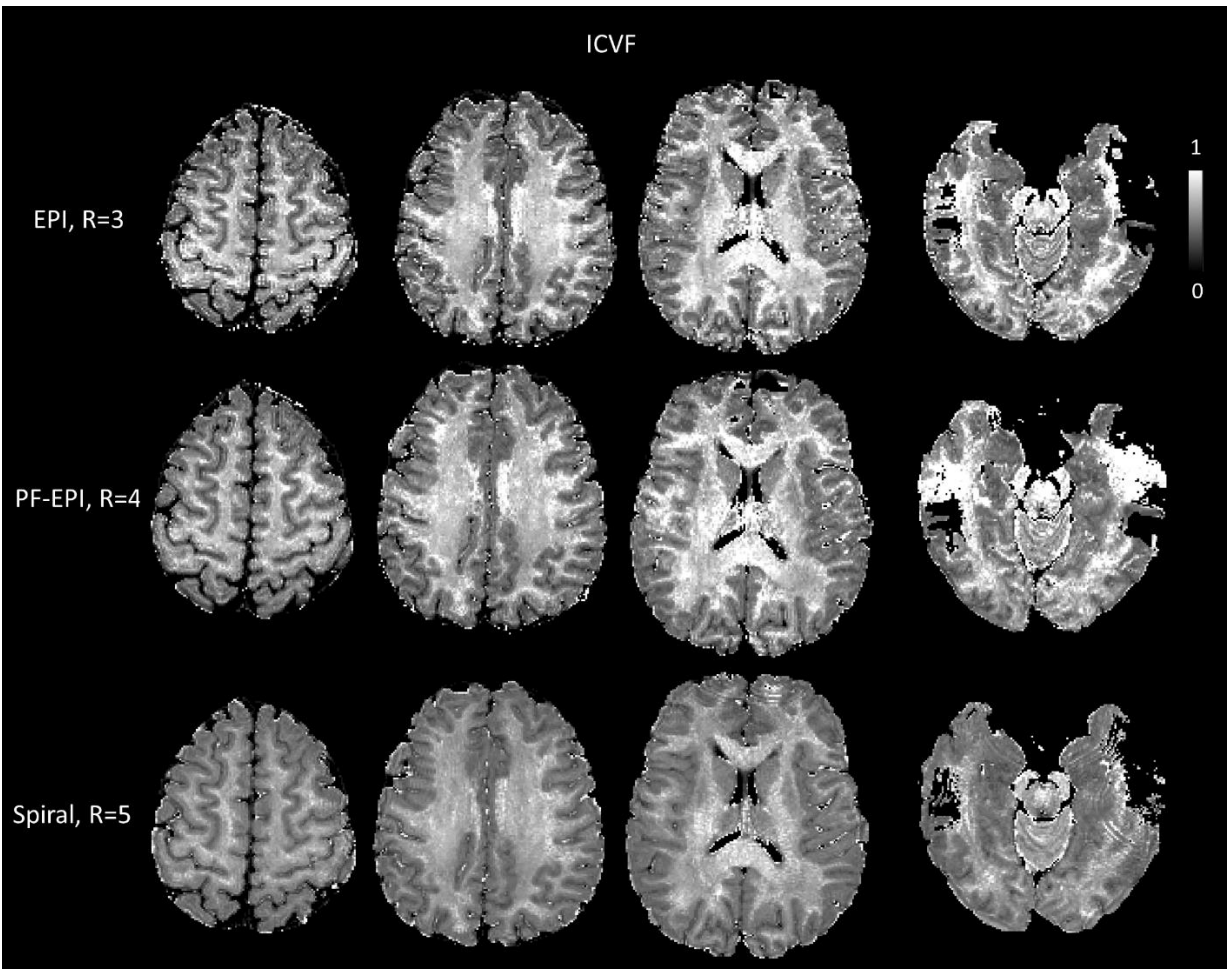

*Figure S13- ICVF maps generated from different trajectories with the same effective resolution. Low SNR of EPI and PF-EPI leads to inaccurate estimation of ICVF. B₁ nonuniformity causes SNR loss in some areas which are clear in the last column.*

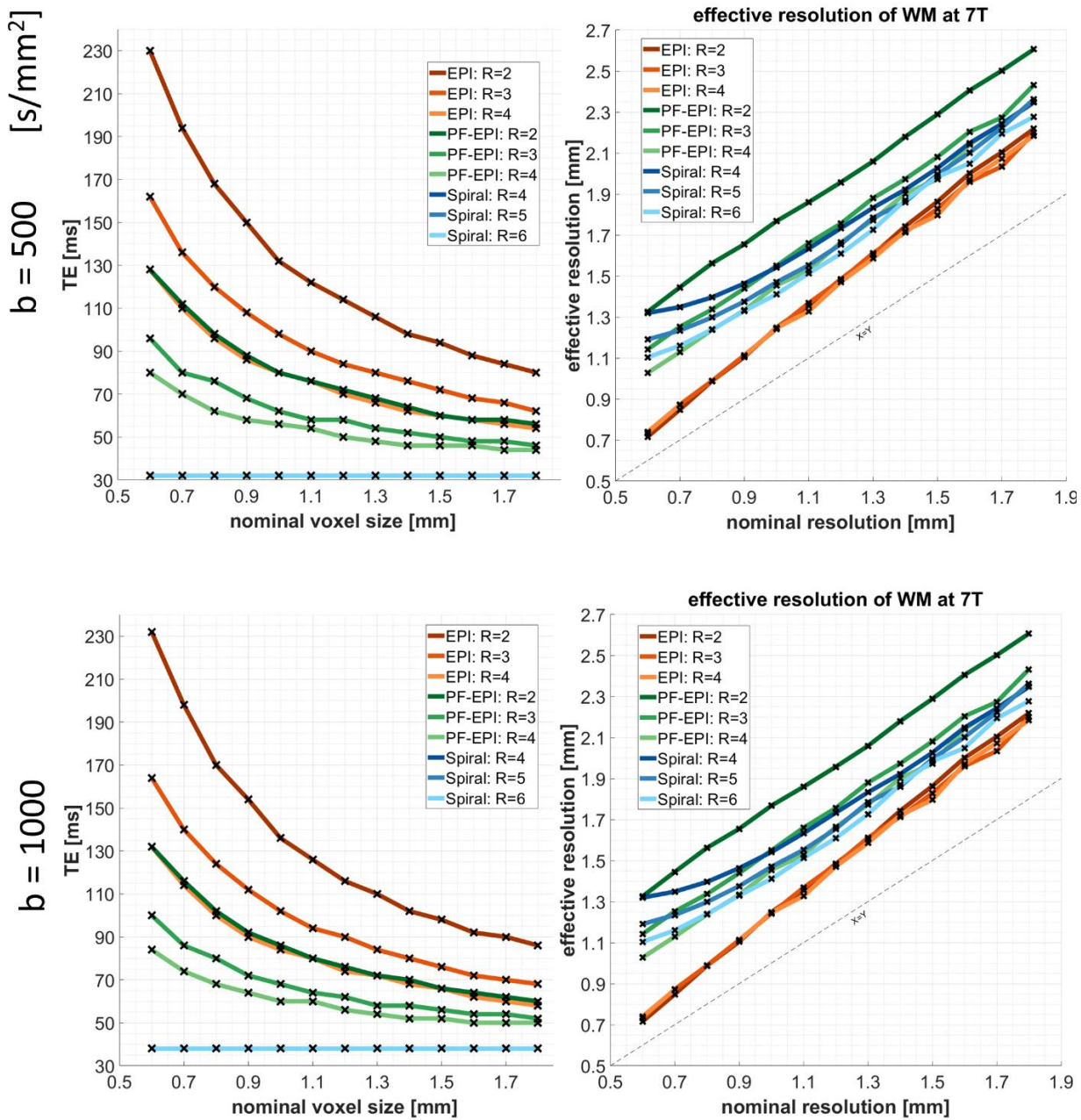

***Figure S14- The echo time and effective resolution as a function of nominal resolution for b-values of 500 and 1000 s/mm². TEs are shorter in b = 500 s/mm², but effective resolution is not affected since the blurring mostly caused by T2\* decay after the echo in a spin-echo sequence.***

**Digital brain phantom simulations**

**Methods**

The PSF analysis above does not include the effects of CG image reconstruction on the image quality. We therefore performed simulations using a digital brain phantom as well. The digital phantom was generated from a segmented T1-weighted image acquired using the MP2RAGE sequence at 7T (Marques et al., 2010) with an isotropic resolution of 1 mm. The WM, GM and cerebrospinal fluid (CSF) compartments were assigned the same relaxation times as for the single-point PSF simulations, and a proton density of 0.55, 0.85 and 1.0 respectively.

The digital phantom was multiplied by the sensitivity maps of the individual channels of the 32-channel Nova coil estimated from a gradient-echo (GRE) scan of the same participant using ESPiRIT (Uecker et al., 2014). T2*-modulated images of the phantom were generated at the time points of the EPI, PF-EPI and spiral trajectories generated for a resolution of 1.5 mm with the same parameters used in the PSF analysis. Points of the k-space were calculated by applying the Fourier transform to the modulated images at the corresponding time points of the trajectories. The under-sampled k-space, coil sensitivity maps and nominal trajectories were then used to reconstruct images using the reconstruction method described in section 2.1. To eliminate differences in image contrast due to echo times, TE was set to 50 ms for all trajectories. This will affect the SNR but will not have any effect on the effective resolution that is investigated. The effects of different trajectories on the resolution were then investigated by qualitatively comparing structural features and line plots.

**Results**

Figure S15 shows simulated spin-echo images created by applying the full image acquisition and reconstruction pipeline on a digital brain phantom. Blurring is visible in the images created using a PF-EPI and spiral trajectory compared to EPI with minimum blurring. By closely looking at magnified regions, the blurring mainly appears along the PE direction in the EPI and PF-EPI images, whereas in images generated using spiral trajectories blurring is spread in all radial directions. Similarly, ringing artifacts in EPI and PF-EPI are mainly affecting voxels in the PE direction, whereas in spirals, voxels in all directions are affected.

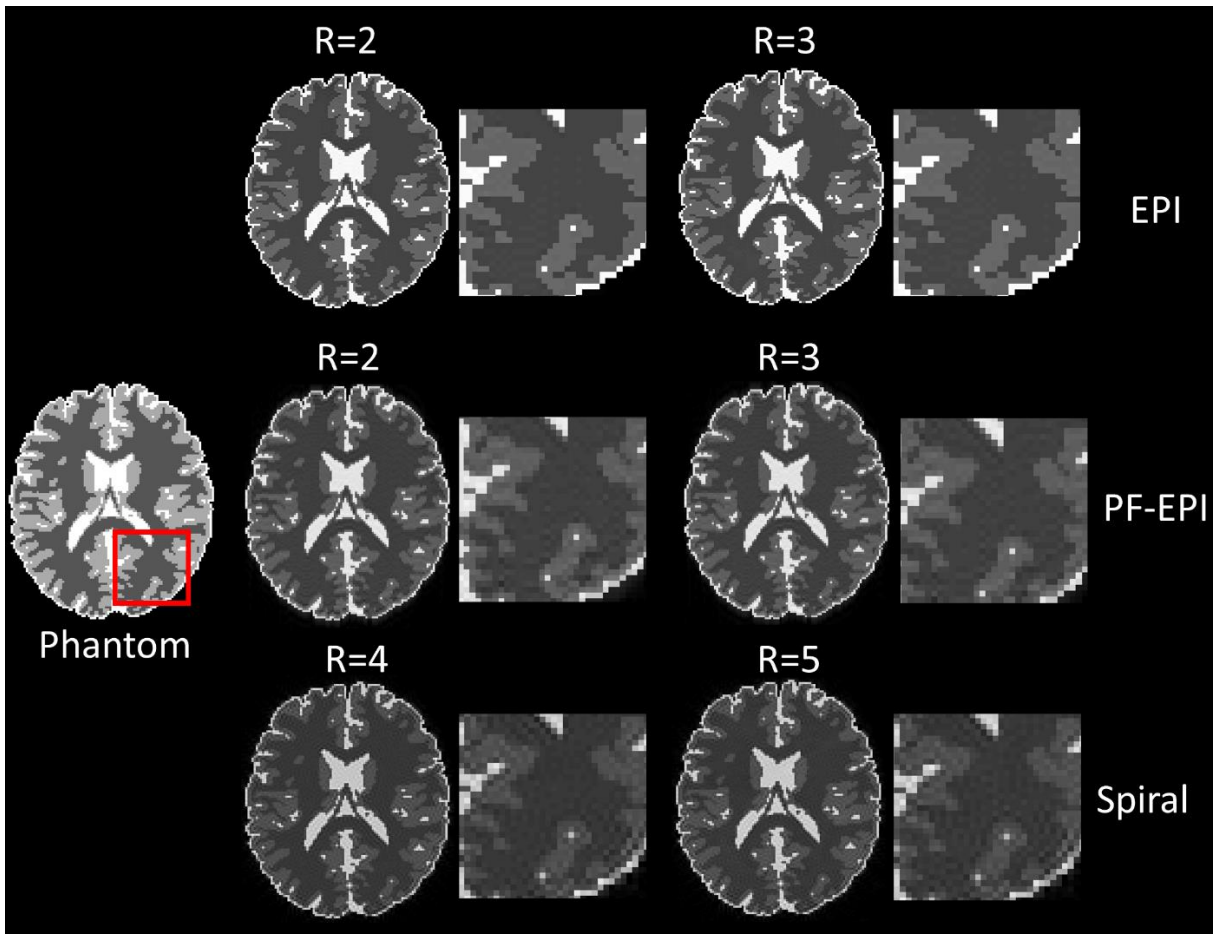

*Figure S15- Digital phantom imaging results. EPI and PF-EPI have higher ringing artifact and blurring in the PE direction, while in spirals they appear in all directions. Blurrier images and more ringing artifacts cause lower specificity and effective resolution.*

Line plots of the reconstructed images in Figure S16 for EPI and PF-EPI with R=2, and spiral R=4 along the anterior-posterior and left-right directions are shown in Figure S16. In the anterior-posterior direction, EPI follows sharp changes clearly, while PF-EPI and spiral smooth out structural details. In the left-right direction, spiral trajectory performs the same as in the other direction, while PF-EPI line plots are similar to EPI due to less blurring in the FE direction.

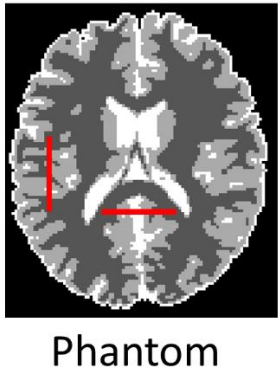

Phantom

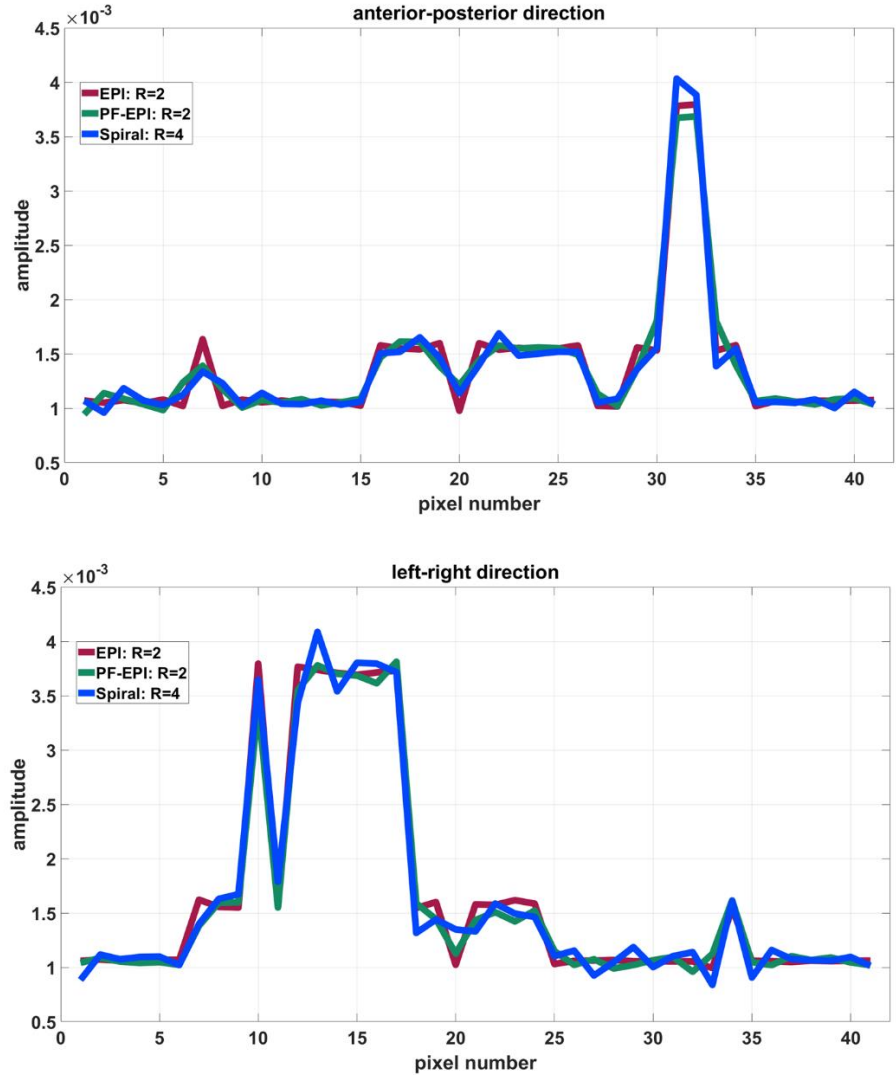

**Figure S16- Line plots from the digital brain phantom images in the phase- and frequency-encode directions.** *The line plots correspond to the red lines in the digital brain image to the left. Spiral and PF-EPI smooth details of the phantom in the anterior-posterior direction. EPI and PF-EPI show similar structural details of the phantom in the left-right direction, while spiral performs similarly to the anterior-posterior direction, losing fine structures.*